# Nanoscale reversal of stable room temperature ferroelectric polarization in organic croconic acid thin films

*Dr. Sambit Mohapatra [a,\*], Dr. Eric Beaurepaire[a,#], Prof. Wolfgang Weber [a], Dr. Martin Bowen [a], Prof. Samy Boukari [a], Dr. Victor Da Costa [a]*

[a] Université de Strasbourg, CNRS, Institut de Physique et Chimie des, Matériaux de Strasbourg, UMR 7504, F-67000 Strasbourg, France

\# Deceased on April 24th, 2018.

E-mail: sambit.mohapatra@ipcms.unistra.fr

victor.dacosta@ipcms.unistra.fr

**Abstract:** It was discovered in 2010 that Croconic Acid, in its crystal form, has the highest polarization among organic ferroelectrics. In the context of eliminating toxic substances from electronic devices, Croconic Acid has a great potential as a sublimable lead-free ferroelectric. However, studies on ferroelectric properties of its thin films are only in their early stages and its capability to be incorporated in nanoscale devices is unknown. In this work, we demonstrate, upon ferroelectric switching at the nanoscale, stable and enduring room temperature polarization with no leakage current in Croconic Acid thin films. We thus show that it is a promising lead-free organic ferroelectric toward integration in nanoscale devices. The challenging switching current and polarization reversal characterization at the nanoscale was done using a unique combination of piezoresponse force microscopy, polarization switching current spectroscopy and the concurrent electromechanical strain response. Indeed, this combination can help to rationalize otherwise asymmetric polarization-voltage data and distorted hysteresis due to current jumps below the background noise, which are statistically washed away in macrojunctions but become prevalent at the nanoscale. These results are valid irrespective of the ferroelectrics' nature, organic or inorganic. Beyond the potential of Croconic Acid as an ecological ferroelectric material in devices, our detection of a clear nanoscopic polarization switching current thus paves the way for a fundamental understanding and technological applications of the polarization reversal mechanism at the nanoscale.

**Main text**

Organic ferroelectric thin films possess great potential for future electronic devices due to certain advantages over their inorganic counterparts: being environmentally benign, less expensive, bio-compatible, easier and simpler to fabricate. As organic material, they are naturally lead-free and thus can be considered in the context of eliminating toxic substances from electronic devices as a lead-free alternative to inorganic ferroelectrics.[1],[2],[3]

As an alternative to inorganic ferroelectric, polymer ferroelectrics are widely studied,[4],[5] but their polarization is significantly smaller than that of standard inorganic ferroelectrics.[6],[7] Ferroelectric polymers cannot be vacuum evaporated and are usually deposited as thin films by spin coating. This limits their incorporation only to multilayer devices where the layers below the polymers can withstand spin coating methods and the necessary post-treatments like thermal annealing.[8],[9],[5] In particular, their deposition on reactive magnetic metal substrates may be problematic due to oxidation of the substrate itself.

However, ferroelectricity was recently discovered in a single component organic material, Croconic Acid (CA), with a polarization value at room temperature of 30 µC/cm$^2$ in its crystal form that rivals that of commonly used inorganic ferroelectrics.[10],[11] Moreover, CA is sublimable such that a thin film deposition under vacuum conditions is possible,[12],[13] which offers the possibility to fabricate various CA-based high-quality multilayer structures.

This high value of polarization is attributed to the unique additive nature of the electronic and protonic contributions to ferroelectricity in the resonance-assisted hydrogen bonded network of molecules.[11] Furthermore, owing to the proton-tautomerism based topological switching mechanism through the π-bond system, polarization reversal takes place at small electric fields,[10] paving the way towards energy efficient low-voltage device operations.

Nevertheless, while several types of macroscopic studies on relatively large crystals and scanning probe microscopy based growth studies on ultra-thin films on metallic surfaces have been performed,[14],[15] studies of the ferroelectric polarization reversal and its kinetics of thin films of CA are only in their early stages. Whether it retains its promising polarization properties down to the nanoscale, toward nanodevice integration, is unknown.

We experimentally demonstrate that CA retains a high polarization even upon switching at the nanoscale in thin films. Before discussing our results, we note that the standard method of characterizing any ferroelectric material is to measure the polarization versus voltage (P-V) hysteresis loop. To do so, the polarization switching current is measured in a capacitor geometry in which the ferroelectric serves as the capacitor dielectric.[16] These switching currents upon polarization switching are usually in an easily measurable range, as the lateral dimensions of the capacitors are typically ~10 μm. However, as the polarization switching mechanism of an ensemble of nanoscopic regions is a statistical process, this hampers access to the ferroelectric properties in individual nanoscopic regions. Furthermore, conduction leakage currents appear through capacitor devices due to conduction hotspots,[17],[18],[19] trapped charges and/or impurities in the dielectric spacer.[20],[21],[22]

Beside macroscopic characterizations, the most widely used technique to characterize local ferroelectric properties is Piezoresponse Force Microscopy (PFM) which can be used to image ferroelectric domains, locally manipulate the polarization state and record local hysteresis loops. However, an interpretation of PFM results in terms of ferroelectricity can be impaired by non-ferroelectric effects such as strain and electrostatic effects due to impurities.[23],[24],[25]

Recently, polarization switching currents were measured at the nanoscale on standard inorganic ferroelectrics such as $BiFeO_3$ (BFO) and $Pb(Zr_{0.2},Ti_{0.8})O_3$ (PZT) that have very high polarization.[26] It is based on the direct detection of polarization switching currents from nanoscopic regions of the nanocapacitors or ferroelectric films with the help of a conductive tip in contact with the top electrode or the sample surface. However, since the polarization switching current, which results from the compensation of polarization charges at the sample surface upon polarization reversal from a nanoscopic region, lies in the ~pA range, it is usually challenging to detect it. Due to large background currents, the measurements were difficult and yielded polarization values of 73.7 $\mu C/cm^2$ for BFO and 119 $\mu C/cm^2$ for PZT.[26] From these results, one may anticipate that it would be even more difficult to study the nanoscale switching currents and polarization reversal of materials with comparatively smaller polarization. To tackle this challenge in CA, we designed a unique set of experiments that combines piezoresponse force microscopy, polarization switching current measurements and concurrent electromechanical strain response.

The PFM characterizations were performed in a UHV microscopy unit at room temperature in contact mode. The ultra-high vacuum environment and in-situ PFM measurements minimize the presence of impurities that would otherwise spoil the analyses of the measured switching currents.

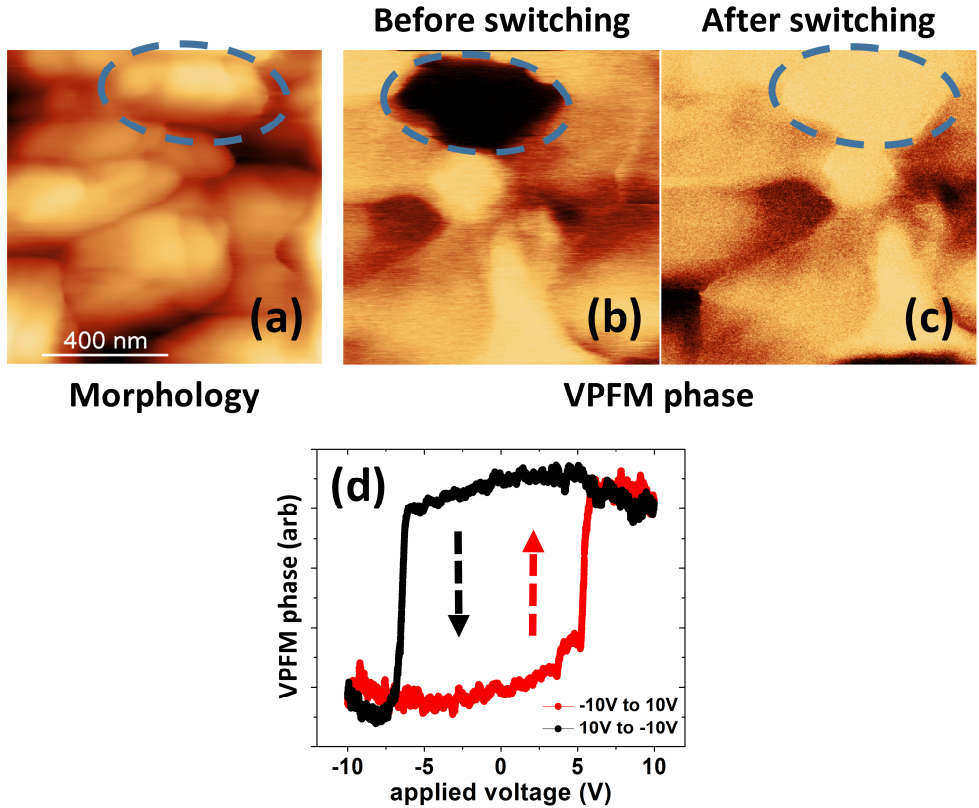

**Figure 1.** PFM domain imaging of a CA (50 nm)/Co (4 nm) sample. (a) shows the surface morphology of the region where the dashed line encircles the nanoscopic region under study. (b) and (c) show the out of plane or vertical polarization component before and after the switching event respectively. (d) shows the characteristic PFM hysteresis loop of the encircled region. Out of plane/vertical phase (VPFM) data is shown (red/black curves) as a function of applied voltage and arrows show the directions of increment of the applied electric field. The voltage increment step was 0.5 mV with a time lapse of 100 µs between two neighboring voltage points. An AC excitation signal of 5 V amplitude and 5.4 kHz frequency was applied during the hysteresis loop measurement. Scaling in the maps is normalized to minimum and maximum values. For (a), the scale goes from 0 to 58 nm.

Domain visualization and hysteresis loop measurements were obtained using our UHV PFM setup. The polarization switching current was collected by a conductive Pt tip in contact with the surface and measured with a high-gain current amplifier stage inside the UHV microscopy unit. The polarization reversal was triggered by applying a voltage ramp to the conductive tip in contact with the nanoscopic region. The switching current measurements were complemented by the synchronous detection of the concurrent piezoelectric nanoscopic strain response. The combination of these three signals characterizes the ferroelectric properties of the material without ambiguity.

We first scan the surface of a 50 nm-thick CA thin film to obtain its morphology (figure 1a) and the map of the ferroelectric domains using the PFM phase (figure 1b, c). To image the ferroelectric domains, an alternating (AC) modulation signal is applied to the tip and its deflection signal is measured with a photodetector and a lock-in amplifier to obtain the piezoresponse-phase signal. The AFM image reveals the presence of nanoscopic regions with a lateral size of several 100's of nm, which are associated with PFM contrast.

Then, we study the local ferroelectric switching characteristics of the encircled nanoscopic region in figure 1a. To do this, an AC excitation/modulation signal along with a DC voltage ramp is applied to the PFM tip in contact with this region. From the PFM contrast differences (figures 1a, b, c) and the obtained hysteresis loop (figure 1d) it is clear that it is possible to switch this nanoscopic region.

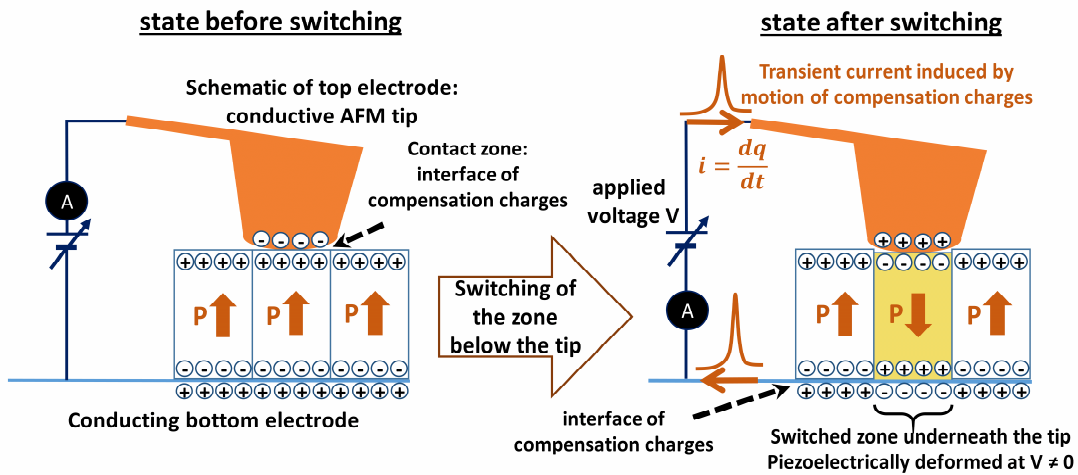

**Figure 2.** Schematics of polarization switching event. The upward polarization of the central zone in the initial state is reversed by the application of voltage via the AFM tip in contact with the zone. The switching results in the flow of charges in the circuit to compensate the new polarization charges that appear at the surfaces of the switched zone after the polarization reversal. This generates current peaks ($i=dq/dt$) whose shape depends on the kinetics of the switching process and the associated compensation process.

The PFM signal alone does not yield the polarization, and it is known that the coercive field depends on the amplitude of the applied AC signal.[27] To get quantitative insight into the reversal, we completed the PFM measurements by detecting polarization switching current and the simultaneous piezoelectric strain response measurements. To measure the polarization switching currents originating from the compensation of the nanoscopic surface polarization charges, we switched off the AC modulation/excitation signal to avoid external periodic perturbations, applied only the DC voltage ramp (-10 V to +10V for switching and then +10 V to -10 V for back-switching) to the tip, and measured the tip current (I) as a function of applied voltage (V). The origin of polarization switching current is schematized in figure 2. The piezoelectric strain response of the nanoscopic region was obtained by directly and simultaneously capturing the real time electromechanical response of the tip-cantilever assembly during the spectroscopic I-V measurements. This gives a direct nanoscopic measurement of the mechanical deformation of the region that occurs synchronously with the polarization reversal process via the inverse piezoelectric effect.[28]

Figure 3a shows a characteristic I-V dataset obtained from the aforementioned region after subtracting a linear background (due to a sample-extrinsic circuit capacitance effect). The noise in our I-V measurement lies in the sub-pA range (rms noise level ~ $2\times10^{-13}$ A). This helps us to detect neatly very small values of switching currents originating from nanoscopic regions of the material. Two sharp switching current peaks can be clearly seen as the bias voltage is swept up, then down. Although these peaks are of similar amplitude, their bias positions are asymmetric. Beside these switching current peaks, we observed no measurable conduction current of any kind (tunnel, leakage, trapped charges…). The absence of bulk-limited conduction currents indicates the absence of any trapped states in the bulk of CA layer.[29],[30] This is in sharp contrast to what is usually observed in inorganic ferroelectric materials,[30] which suffer from conduction leakage currents due to trapped defects,[22] oxygen vacancies[31] etc. at similar film thicknesses.

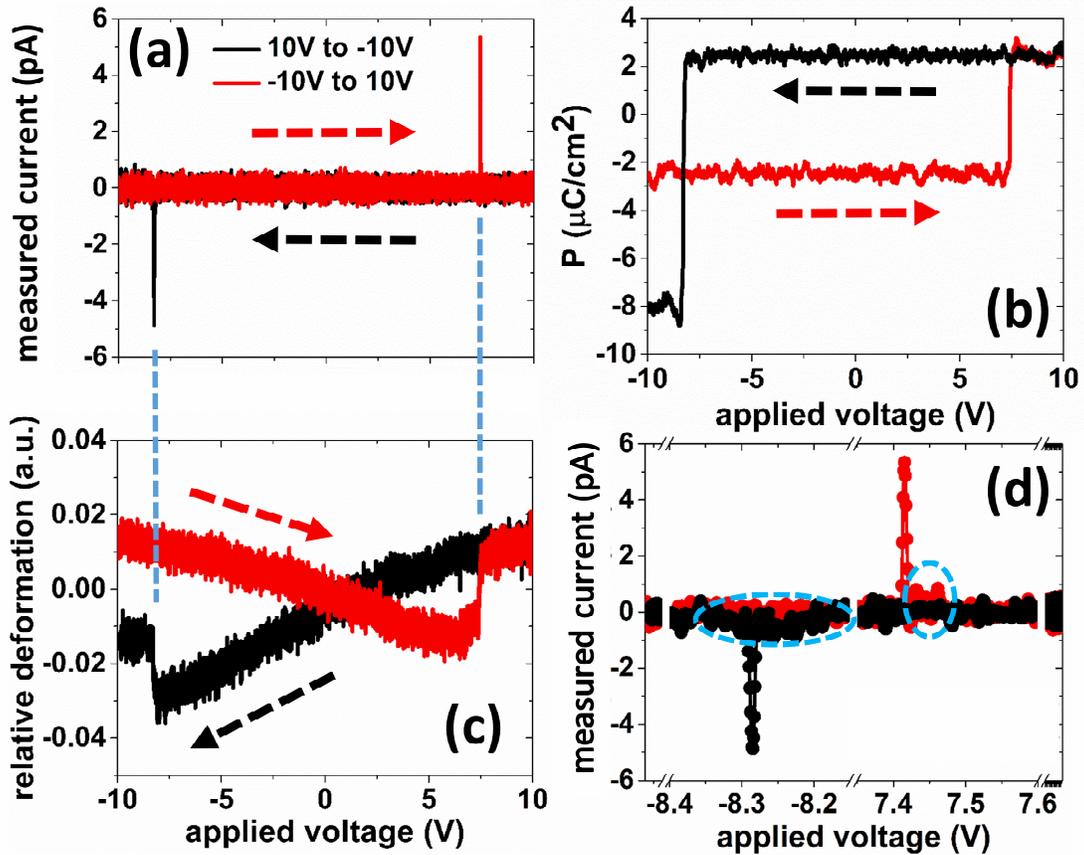

**Figure 3.** Switching current measurements at region 1. (a) shows the observed I-V curve with the characteristic current peaks corresponding to polarization switching events. (b) shows the estimated P-V curve, where P is the polarization charge density in µC/cm$^2$. The strain response (relative electromechanical deformation) of the nanoscopic region to the switching event is shown in (c). The sudden jumps in the deformation are characteristic of polarization switching. The synchronization between the currents peaks (dashed blue lines) in (a) and jumps in (c) proves that the current peaks in (a) are indeed due to switching events. (d) shows a close up view of the I-V plot near the switching current peaks. The dashed circles encircle the regions of tiny but existent current fluctuations near the current peaks. No AC modulation/excitation signal was applied during measurement.

It is evident that the major switch/jump in the deformation (figure 3c) and the current peaks in figure 3a occur at the same voltage; the simultaneity in the occurrence of the two events confirms a real polarization reversal giving rise to the current peaks. But when compared to the PFM loop (figure 1), it is obvious that the reversal with only the DC signal occurs at higher bias. This is consistent with the observation that the coercive field depends on the AC amplitude value.[27] PFM is thus convenient to obtain a loop, however, one has to be cautious with the measured coercive field. Indeed, since the coercive field is defined as the strength of the electric field at which the polarization of the ferroelectric capacitor disappears, it does not necessarily correspond to the positions of the switching current maxima. Thus, to get the coercive field and the polarization, one has to calculate the Polarization P-V hysteresis loop.

The bias dependence of P is obtained by integrating the measured current of figure 3a over time (each voltage point corresponds to 100 µs), divided by the switched area. A switched area of ~0.103 µm$^2$ was determined from figure 1 (a: before switching and b: after switching). The as-calculated P-V loop is shown in figure 3b. Strikingly, even though the height of the switching current peaks in figure 3a is almost the same, the loop is not closed and there is a significant difference between the jump in polarization values for the two directions of the applied voltage. We estimate the value of P by taking the half of the jump or change in polarization for a given sign of applied voltage. Thus, polarization values of 5 µC/cm$^2$ and 2 µC/cm$^2$ and coercive fields of -8.24 V and 7.41 V are obtained at negative and positive bias, respectively. The positions of the major peaks in figure 3a and figure 3c are thus close to the coercive field, but not exactly the same.

This asymmetry in the polarization values on both sides of applied voltage can reflect how the switching current amplitude depends not only on the area of the switched region but also on the kinetics of the switching process in that region. As the polarization is proportional to the time integration of the measured instantaneous current (area under the current peak), switching kinetics greatly affects the estimation of its magnitude. Although in figure 3a the two current peaks appear to be similar, a close-up (figure 3d) reveals that the peaks are not symmetric: their shapes are different and the peak at positive bias spans over ~1300 µs, whereas this time span is ~1800 µs at negative bias. This asymmetry in the temporal spread of the switching current peaks translates into an asymmetry in the estimated P-V hysteresis loop.

This can be explained if we suppose that a significant number of switching events result in currents that go undetected due to the presence of the low frequency background noise and the limited detection capability of our instrument. Such a scenario is possible if the net reversal event consists of multiple steps, many of which produce very small switching currents that are buried within the noise level. In the strain response (See supplementary figure S2), such a multi-stepped switching appears as a gradual jump (positive bias) as opposed to a sharp jump (negative bias) for a single step switching event. A more significant portion of the switching process is captured by the current peak at the negative bias, owing to the sharpness or single stepped nature of the reversal, resulting in a larger spread of the current peak. This is the reason why the estimated P-V loop in figure 3b shows a larger change in polarization on the negative side than on the positive side.

Thus, it is evident that the asymmetry in switching kinetics of the polarization reversal process results in an asymmetry of the polarization densities in the P-V loops. The maximum value of ~5 µC/cm$^2$ that we obtained for the saturation polarization at negative voltage is a lower limit of the saturation polarization in our measurement.

Given the stochastic character and asymmetry of the reversal and its kinetics, one has to examine how the P-V loops and thus the estimated polarization and coercive field changes from switching to switching. We thus repeated the spectroscopic measurements eleven times on the same nanoscopic region, keeping all parameters unchanged (see supplementary figure S3). The height of the current peaks and their positions on the voltage axis are shown in a scatter plot (figure 4a). It is clear that the peaks appear at random voltage positions. Their heights and positions are asymmetric with respect to the applied electric field direction for all reversal events. Two groups of points are visible in the scatter plot; one for negative voltages and the other for positive voltages. The corresponding means and standard deviations for the peak positions-peak heights are (-8.3±0.3 V; -3.7±0.9 pA) and (7.0±0.7 V; 4.2±1.6 pA), respectively.

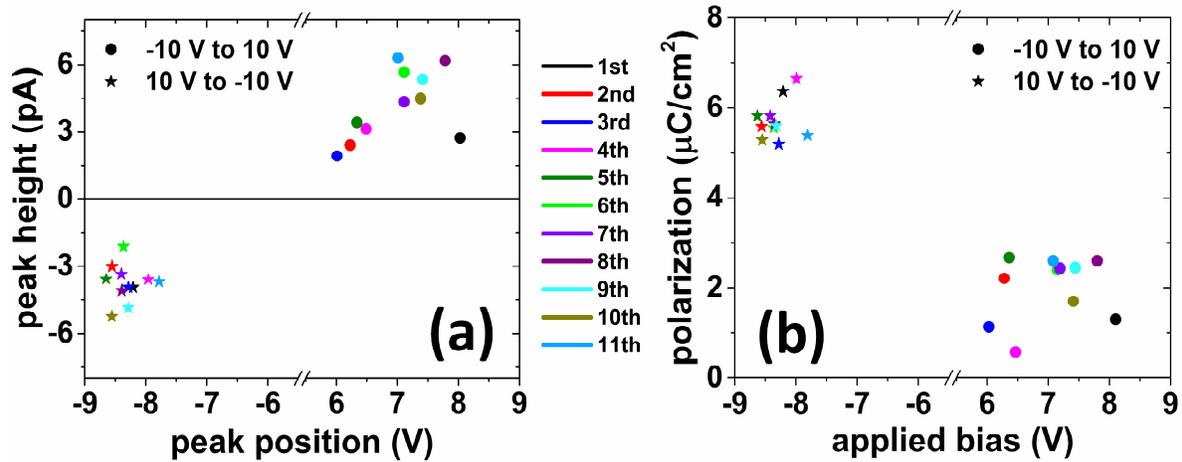

**Figure 4.** Stochastic nature of polarization switching of the region shown in figure 1. (a) shows a scatter plot of the amplitude of the switching current maxima in the I-V curves (supplementary figure S3) with respect to the applied voltage position of the current peaks. (b) shows the corresponding scatter plot of the estimated polarization values as a function of the applied voltage, where polarization values are estimated separately for each direction of the applied voltage. Dots and stars represent the negative to positive and positive to negative directions of applied voltage, respectively. The ensemble of repeated I-V measurements were performed on the same nanoscopic region with all experimental parameters unchanged. Each color represents one measurement. The random behavior of the distribution depicts the stochastic nature of the polarization switching processes.

A scatter plot of the estimated polarization values corresponding to each spectroscopic switching current measurements of figure 4a is shown in figure 4b. Quantitatively, the means and standard deviations for the voltage positions and polarizations, on negative and positive sides, respectively, are (-8.3±0.2 V; 5.7±0.4 $\mu C/cm^2$) and (7.0±0.7 V; 2.0±0.7 $\mu C/cm^2$). The larger dispersion of the coercive fields and polarizations on the positive side indicate an asymmetry in the reversal process with respect to the direction of the applied field. The largest estimated polarization of 6.6 µC/cm$^2$, albeit the lower limit, which was obtained for the fourth reversal is ~78% smaller than the maximum reported value of saturation polarization for CA crystals but is still on the higher end among room temperature organic ferroelectric materials.[6]

While the high value of polarization in CA has been reported only for macroscopic crystals, structural differences[32],[33],[13] may result in a reduced value of saturation polarization in our thin films. Further, a canted alignment of the polarization vector[13] in the nanoscopic regions of polycrystalline thin films restricts the estimation of the net saturation polarization (vector sum of the out-of-plane and in-plane components), unless the canting angle is exactly determined, which requires the information on the entire piezoresponse signal (PFM phase and PFM amplitude).[34],[35] Thus, our reported 6.6 µC/cm$^2$ maximum value for the CA polarization represent only the out-of-plane component of the polarization; the magnitude of the polarization vector can be much higher.

Besides, we also observe in figure 4b that the loops are also biased toward negative values, confirming what was observed on the single loop of figure 1d. This shift is probably due to an internal bias field whose magnitude and distribution depend on the composition, microstructure and on the thermal and electrical history of the system.[36] Interestingly, although for a single experiment the current peak maxima and the coercive fields do not necessarily coincide in our eleven experiments, the mean values for their positions is exactly the same. Again, we see that the switching processes take place at different applied biases depending on the presence or absence of the AC excitation signal, as can be seen on the coercive field values with AC excitation ($H_c$ ~5 V, figure 1d) and without ($H_c$ ~7 V figure 3 and figure 4).[27] Note that here also, due to the polarization canting angle, we could expect to observe a much smaller coercive field for an out-of-plane polarization.

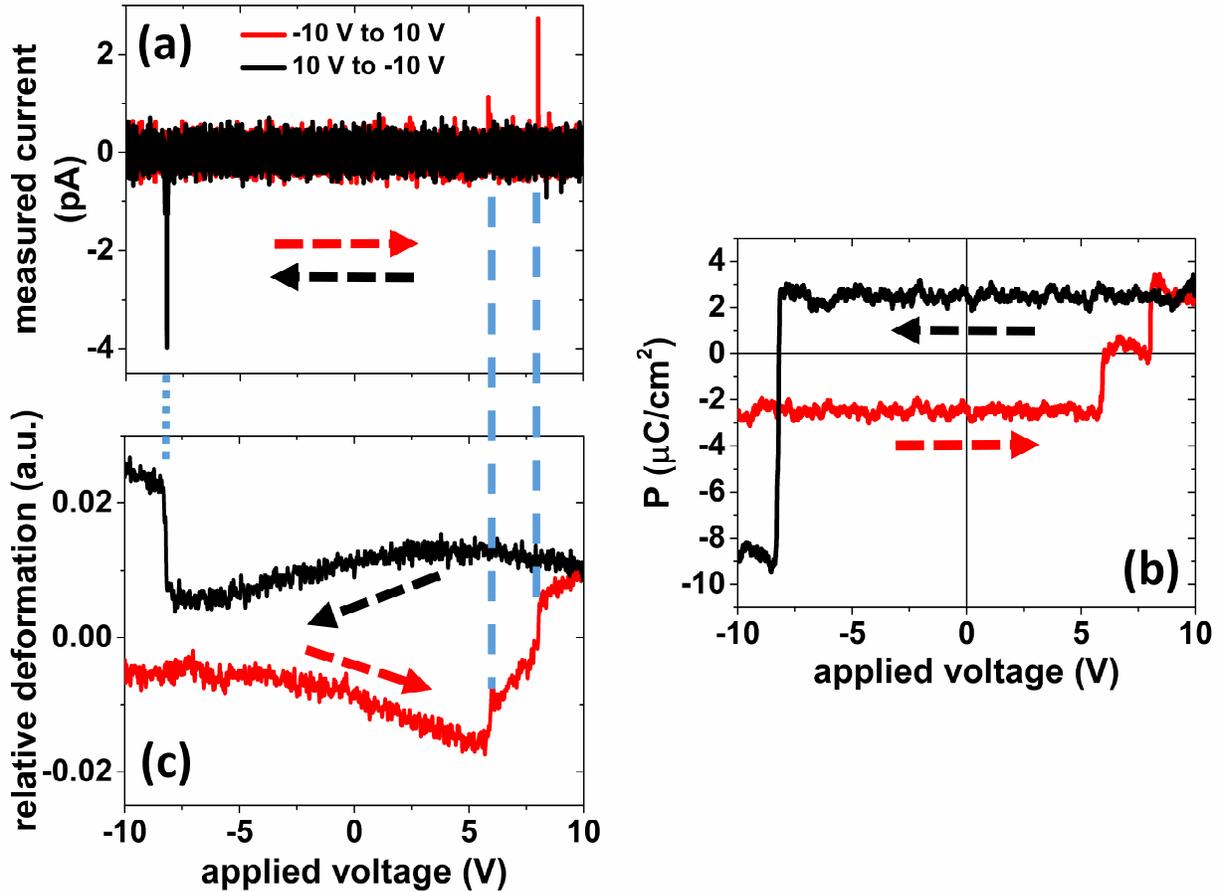

**Figure 5.** The first measurement of switching current spectroscopy in the stochastic ensemble of measurements of figure 4. (a), (b) and (c) show the I-V, estimated P-V hysteresis loop and the piezoelectric stain response of the spectroscopic measurement, respectively. The dashed arrows indicate the direction of application of the DC ramp voltage. The simultaneity in the occurrence of switching current peaks and strain response jumps, as indicated by the dashed lines, confirms real switching events. Multiple steps in P-V hysteresis loop in (b) correspond to the multiple switching current peaks in (a).

We emphasize that the dispersion in our data is due to the stochastic character of the reversal, which is averaged at the macroscopic level but is particularly prominent here at the nanoscale. This is clear if we look at a particular I-V spectroscopy plot, say the first measurement (black curve) in figure 4a, as presented in figure 5. We see that the positive side of the applied voltage has two steps in the polarization hysteresis loop (figure 5b) corresponding to the switching current peaks (figure 5a) which are also in perfect synchronization with the two jumps in the strain response curve (figure 5c). However, on the negative side, only one step is present in the hysteresis loop (figure 5a) corresponding to the single switching current peak (figure 5a). The fact that the step-wise reversal behavior is more evident (figure 5) and the polarization values are always smaller on one side (positive side) of the applied voltage (figure 4b) again hints at an asymmetry in the reversal kinetics. This is also evident in the strain response data (see supplementary figure S4 and S5) for the ensemble of spectroscopic measurements corresponding to figure 4.

The asymmetry in individual spectroscopic measurements can be even more pronounced on several other nanoscopic regions (figure 6 represents one such region) of the same sample. Indeed, we observe multiple switching current peaks along one direction of the applied electric field and a smaller or almost non-existent peak along the other direction (figure 6a), while the signatures of the switching events are clearly visible along both directions as jumps in the strain response curve (figure 6b). The presence of current peaks corresponds, in the strain response, to abrupt jumps and their absence to a very gradual jump. Figure 6c shows an expanded view near the jumps in the strain response curves, focusing on the switching region (inside the dashed line in figure 6b). It is clear that the reversal, when applying voltages from 10 V to -10 V, is gradual, whereas from -10 V to 10 V it consists mainly of two sharp steps (blue arrows in figure 6c) that correspond to the two peaks in the measured current (figure 6a). We thus have clear evidence that the switching current corresponding to the gradual reversal can go entirely undetected due to the background noise present in the data. This is the extreme case of the asymmetric switching kinetics in terms of the applied voltage direction, where an attempt to deduce the P-V hysteresis loop will result in a complete loss of polarization in one direction (here negative voltage side) of the applied voltage, making it impossible to obtain a closed hysteresis loop.

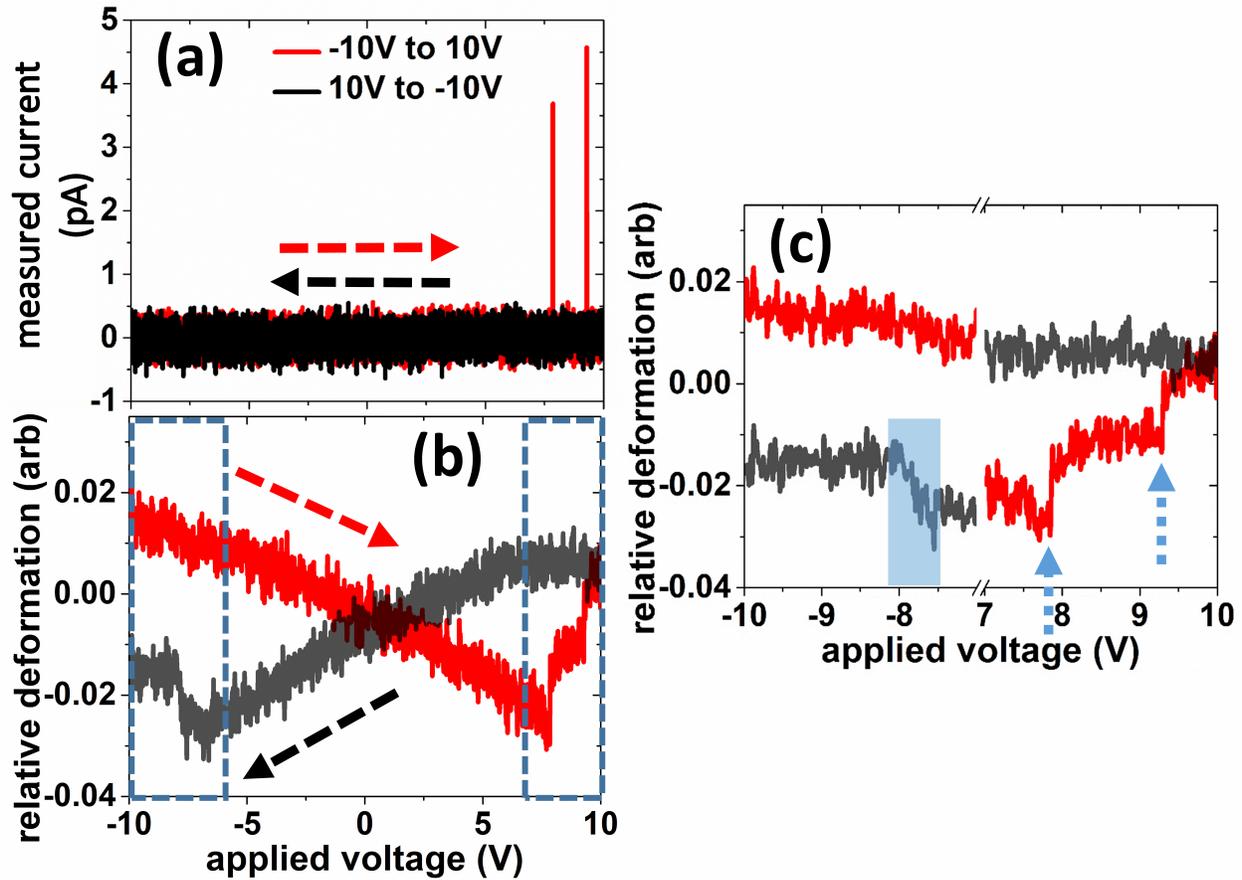

**Figure 6.** Spectroscopic measurements on another nanoscopic region of the same sample. (a) presents the measured I-V curve. Two switching current peaks are present on one side (red), whereas no peak appears on the other side of applied bias (black). (b) shows the corresponding strain response, i.e. the relative electromechanical deformation during the spectroscopic measurement. Presence of jumps in (b) on both sides of applied voltage is an indication of successful polarization switching along both directions of applied field. The absence of switching current peaks on the negative bias side in (a) is attributed to the gradual switching process signified by a smooth jump on the negative side in (b). (c) shows an expanded view of regions marked with dashed rectangles in (b) showing the two sharp abrupt jumps on the positive bias side (dashed arrows) and a smooth jump on the negative bias side (blue rectangular shade).

To conclude, we have demonstrated the stable and enduring polarization at the nanoscale, with no leakage current, upon electrical switching of thin films of Croconic Acid on Co. This makes CA a promising organic ferroelectric for integration into environmentally friendly nanoscale capacitor devices that comprise a nanoscopic region, e.g. a single grain, using resist-nanopatterning or nanobead[37] processing. Our results were obtained by a unique combination of PFM, nanoscopic polarization switching current and piezoelectric strain response measurements that is very pertinent when studying any ferroelectric at the nanoscale prior to device fabrication. Indeed, this combination can help to rationalize otherwise asymmetric polarization-voltage data and distorted hysteresis due to current jumps below the background noise, which are statistically washed away in macrojunctions but become prevalent at the nanoscale. The stochastic reversal that is observed at the nanoscale suggests caution when assessing the projected capabilities of a ferroelectric memory element using a limited dataset. Our detection of a clear nanoscopic polarization switching current thus paves the way for a fundamental understanding and technological applications of the polarization reversal mechanism at the nanoscale.[38],[39],[40]

**Experimental Section**

CA films with a thickness of 50 nm were grown on 4 nm of cobalt at room temperature under ultra-high vacuum (UHV) conditions with a base pressure of ~ $5 \times 10^{-10}$ mbar. Co was deposited by Molecular Beam Epitaxy on mica/Au substrates at a deposition pressure of around $1 \times 10^{-9}$ mbar and CA was deposited in-situ at a growth rate of 0.5 nm/min and at a deposition pressure of around $1 \times 10^{-8}$ mbar. The cell temperature was 125 °C for the CA deposition. Such growth conditions result in a quasi-continuous film with nanoscopic regions, typically 300 nm in diameter.[13] The as-prepared samples were then in-situ transferred to the UHV scanning probe microscopy system with a working pressure of ~ $2 \times 10^{-10}$ mbar.

*Dr. Sambit Mohapatra [a*], Dr. Eric Beaurepaire[a#], Prof. Wolfgang Weber [a], Dr. Martin Bowen [a], Prof. Samy Boukari [a], Dr. Victor Da Costa [a]*

[a] Université de Strasbourg, CNRS, Institut de Physique et Chimie des, Matériaux de Strasbourg, UMR 7504, F-67000 Strasbourg, France

# Deceased on April 24th, 2018.

E-mail: victor.dacosta@ipcms.unistra.fr

sambit.mohapatra@ipcms.unistra.fr

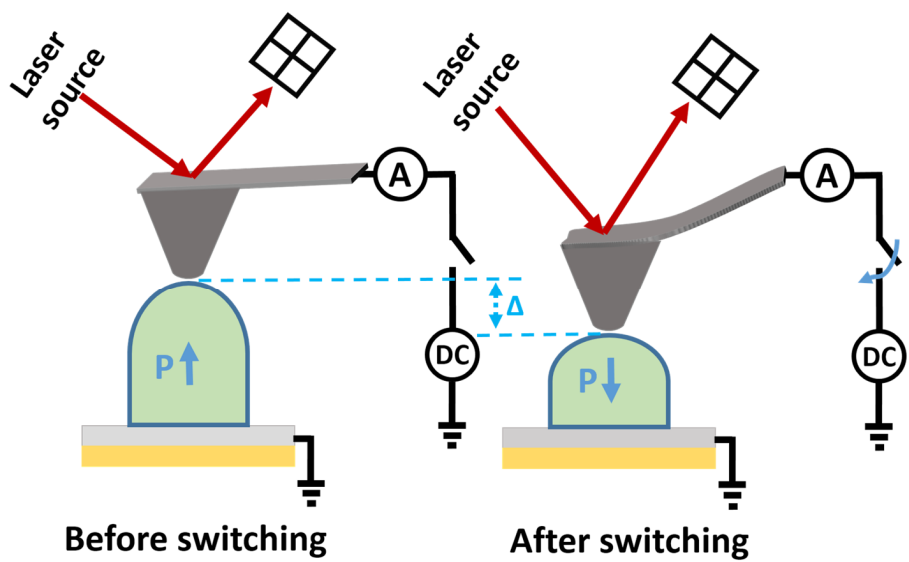

**Figure S1.** Schematics of the origin of piezoelectric electromechanical strain response during a ferroelectric switching event. The change in the deflection of the laser beam due to the polarization reversal induced cantilever deformation is captured by the position sensitive crossed photodetector. The cantilever deformation is proportional to the piezoelectric deformation of the ferroelectric region.

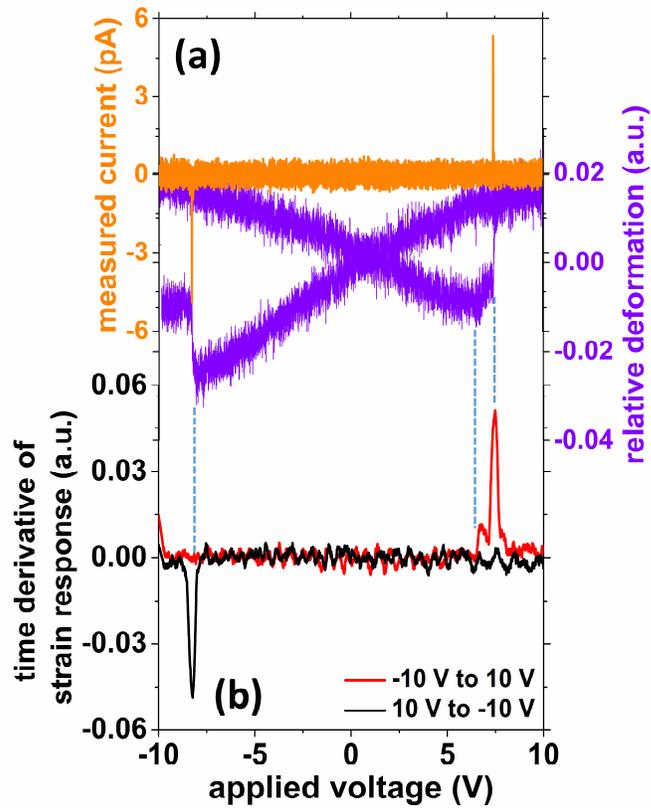

**Figure S2.** Detailed view of the measured current and the strain response corresponding to figure 3. (a) shows the measured current (orange curve) superposed on the strain response (blue curve). (b) shows the first order time derivative of the strain response curve smoothened by Savitzky-Golay filter with points of window = 500. Strain response curve in (a) is purposefully shifted by an appropriate magnitude for a better comparison with the current peaks. Dashed lines indicate the position (negative bias) and spread (positive bias) of the jump on the strain response curve.

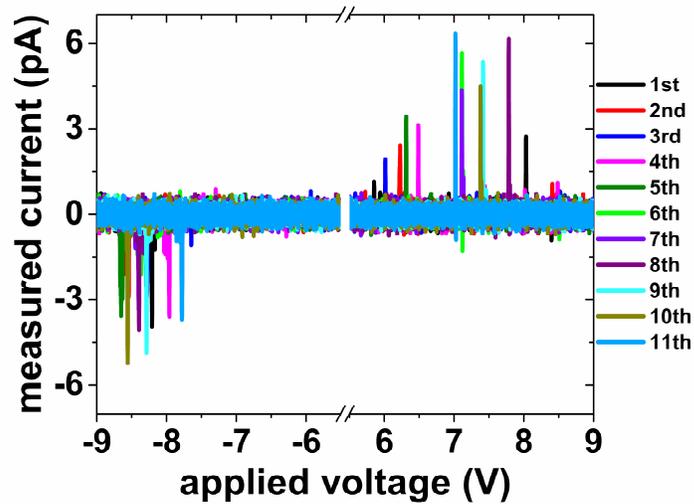

**Figure S3.** Ensemble of I-V spectroscopic curves corresponding to figure 4. Repeated I-V spectroscopies were performed on the same nanoscopic region with a time interval of 100 µs between each two successive measurements. Each color represents one spectroscopic measurement as indicated by the corresponding index.

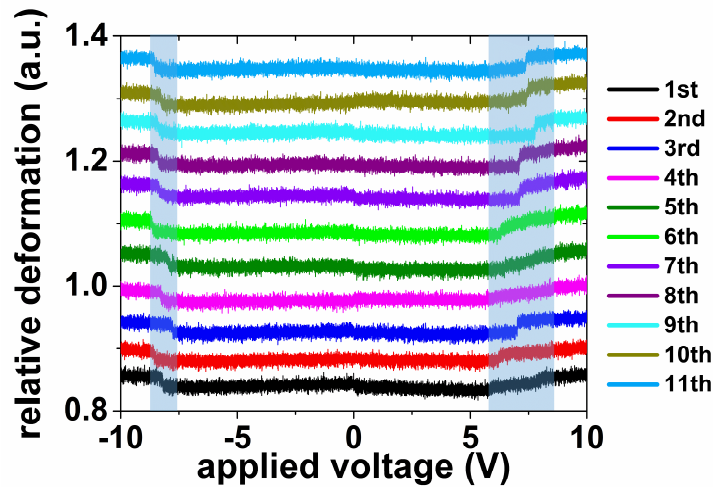

**Figure S4.** Strain response curves of switching spectroscopies of figure 4. For the sake of clarity, only the branches that contain the polarization switching related jumps are shown for each of the curves. Colors represent the sequence of individual spectroscopic measurements. The curves are shifted by appropriate magnitude for clarity. The blue shades show the ranges of applied voltage over which jumps on the deformation curves take place on either side of the applied voltage.

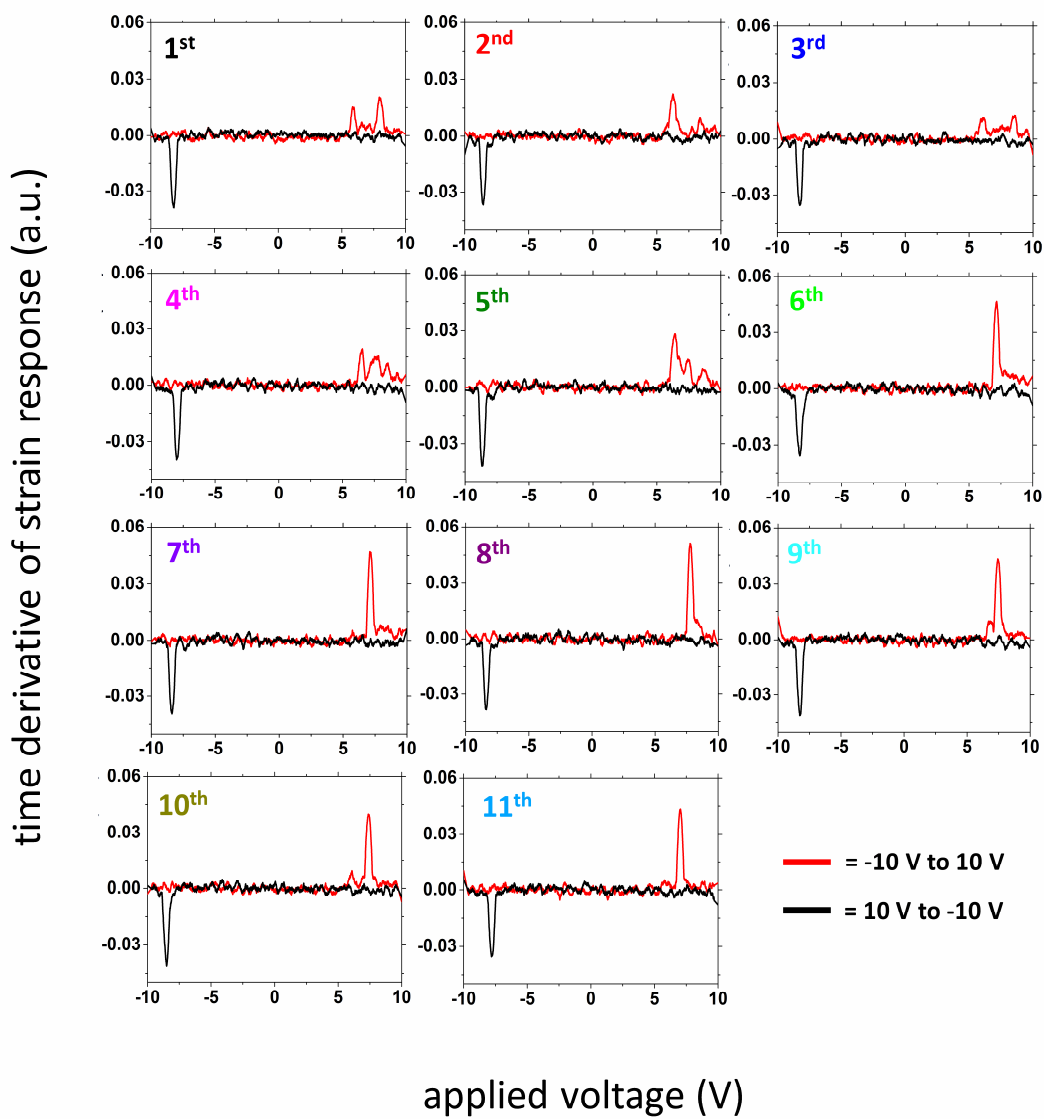

**Figure S5.** Time derivative plots of strain response ensemble corresponding to the spectroscopic I-V measurements of figure 4. Curves are smoothened by Savitzky-Golay filter with points of window = 600. The colored indices correspond to the sequence of individual measurements. The correct way to get information on the reversal kinetics would be to study the spread of the peaks, not the heights, for the magnitude of deformation is not known to be constant for the entire ensemble of measurements.

Figure S4 presents the individual strain response curves of all the 11 measurements corresponding to figure 4 in the main text, where we see that for many of the curves on the positive side of the applied voltage, no sharp jump is visible on the strain response curves; rather a very gradual jump is evident. On the negative side, however, the jump on every single strain response curve is sharp. This is consistent with the multi-stepped or gradual reversal kinetics on the positive side of the applied voltage.

Polarization switching current is an important measurement parameter when it comes to the characterization of ferroelectric materials. While at the macroscale the switching current is widely used to record hysteresis loops and measure the polarization, at the nanoscale however, the switching currents can evidence fast polarization reversal events, where gradual reversals can be missed easily due to the presence of background noise. The polarization values deduced from the switching currents may be significantly biased due to background noise and instrument limitations and thus result in a severe distortion of the hysteresis loops. The shape of the hysteresis loop measured by PFM phase is not affected by these problems but depends on the AC signal applied to the tip. In any case, the reversal at the nanoscale is, in essence, stochastic in terms of the reversal kinetics and the applied voltage at which it takes place. Thus, one has to be cautious before deducing material properties from a single or a limited number of measurements. For a ferroelectric memory element for instance, the voltage required to reverse the polarization with a given probability must be determined by numerous trials.

**Acknowledgements**

We acknowledge the support of funding for the project from the agencies, ANR and DFG (ORINSPIN- ANR-16-CE92-005-01). We acknowledge the efforts of Mr. Jacek Arabski and Mr. Guy Schmerber from IPCMS, CNRS-University of Strasbourg, France for their assistance during the work of the project. We are thankful to Dr. Salia Cherifi-Hertel of IPCMS, CNRS-University of Strasbourg, France for useful discussions.